\documentclass[twocolumn]{aastex62}

\graphicspath{{./}{figures/}}

\shorttitle{Gravity, rotation and magnetic field}
\shortauthors{Beuther et al.}

\begin{document}

\title{Gravity and rotation drag the magnetic field in high-mass star formation}

\correspondingauthor{Henrik Beuther}
\email{beuther@mpia.de}

\author[0000-0002-1700-090X]{Henrik Beuther}
\affil{Max Planck Institute for Astronomy, K\"onigstuhl 17, 69117 Heidelberg, Germany}

\author[0000-0002-0294-4465]{Juan D.~Soler}
\affil{Max Planck Institute for Astronomy, K\"onigstuhl 17, 69117 Heidelberg, Germany}

\author[0000-0002-8115-8437]{Hendrik Linz}
\affil{Max Planck Institute for Astronomy, K\"onigstuhl 17, 69117 Heidelberg, Germany}

\author[0000-0002-1493-300X]{Thomas Henning}
\affil{Max Planck Institute for Astronomy, K\"onigstuhl 17, 69117 Heidelberg, Germany}

\author[0000-0002-8120-1765]{Caroline Gieser}
\affil{Max Planck Institute for Astronomy, K\"onigstuhl 17, 69117 Heidelberg, Germany}

\author[0000-0003-2309-8963]{Rolf Kuiper}
\affiliation{Institute of Astronomy and Astrophysics, University of T\"ubingen, Auf der Morgenstelle 10, D-72076 T\"ubingen, Germany}

\author[0000-0002-2700-9916]{Wouter Vlemmings}
\affiliation{Department of Earth and Space Sciences Chalmers University of Technology, Onsala Space Observatory, 439 92, Onsala, Sweden}
  
\author[0000-0002-0472-7202]{Patrick Hennebelle}
\affiliation{AIM, CEA, CNRS, Universite Paris-Saclay, Universite Paris Diderot, Sorbonne Paris Cite, 91191 Gif-sur-Yvette, France}
  
\author[0000-0002-4707-8409]{Siyi Feng}
\affiliation{Chinese Academy of Sciences Key Laboratory of FAST, National Astronomical Observatory of China, Datun Road 20, Chaoyang, Bejing, 100012, P.~R.~China}
  
\author[0000-0002-0820-1814]{Rowan Smith}
\affiliation{Jodrell Bank Centre for Astrophysics, Department of Physics and Astronomy, University of Manchester, Oxford Road, Manchester M13 9PL, UK}
  
\author[0000-0003-4037-5248]{Aida Ahmadi}
\affiliation{Leiden University, Niels Bohrweg 2, 2333 CA Leiden, Netherlands}
  
\begin{abstract}

  The formation of hot stars out of the cold interstellar medium lies
  at the heart of astrophysical research. Understanding the importance
  of magnetic fields during star formation remains a major
  challenge. With the advent of the Atacama Large Millimeter Array,
  the potential to study magnetic fields by polarization observations
  has tremendously progressed. However, the major question remains how
  much magnetic fields shape the star formation process or whether
  gravity is largely dominating. Here, we show that for the high-mass
  star-forming region G327.3 the magnetic field morphology appears to
  be dominantly shaped by the gravitational contraction of the central
  massive gas core where the star formation proceeds. We find that in
  the outer parts of the region, the magnetic field is directed toward
  the gravitational center of the region. Filamentary structures
  feeding the central core exhibit U-shaped magnetic field
  morphologies directed toward the gravitational center as well, again
  showing the gravitational drag toward the center. The inner part
  then shows rotational signatures, potentially associated with an
  embedded disk, and there the magnetic field morphology appears to be
  rotationally dominated. Hence, our results demonstrate that for this
  region gravity and rotation are dominating the dynamics and shaping
  the magnetic field morphology.

\end{abstract}


\section{Introduction} \label{sec:intro}

How important are magnetic fields during the formation of dense
molecular clouds and the parallel/subsequent star formation processes?
While some works have stressed the importance of magnetic fields
during cloud formation and core collapse (e.g.,
\citealt{mouschovias1979,commercon2011,tan2013,tassis2014,hennebelle2018})
other groups favor scenarios where turbulence and/or gravity are the
dominant physical processes (e.g.,
\citealt{padoan2002,maclow2004,gomez2014,padoan2017,vazquez2019}).
Studies on cloud scales ($\sim$10\,pc) show clear signatures of the
importance of magnetic fields to shape the structure of the
interstellar medium (ISM, e.g.,
\citealt{soler2016,soler2019b,fissel2019}), whereas the situation is
far less clear on sub-pc scales of individual star-forming
regions. Hourglass-like magnetic field morphologies, that are
interpreted as indicative for a tight coupling between the magnetic
field and the dense gas, were observed for several regions (e.g.,
\citealt{rao1998,girart2006,girart2009}). However, for many other
sources the morphologies are less conclusive (e.g.,
\citealt{zhang2014b,hull2014,koch2014,koch2018}). While early
observations of some high-mass star-forming regions indicated that
turbulent energies may be equal to or dominate over magnetic energies
(e.g., \citealt{beuther2010c,girart2013}), other studies found regions
with low turbulent-to-magnetic energy ratios (e.g.,
\citealt{girart2009,beuther2018,dallolio2019}). Whether weak or strong
magnetic fields are typical in star formation is an ongoing debate
(e.g., \citealt{crutcher2010,li2015,pillai2015}). For a summary of the
current state of research based on interferometric polarization
studies of star-forming regions from low- to high-mass stars we refer
to the recent review by \citet{hull2019}.

\begin{figure*}[ht]
\includegraphics[width=\linewidth]{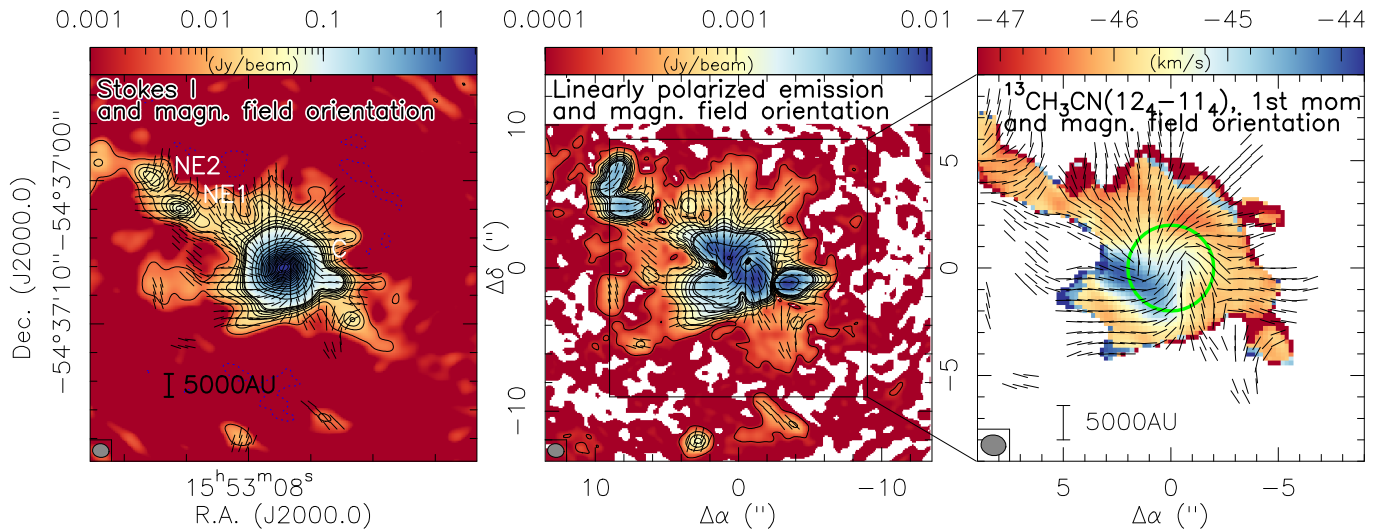}
\caption{Compilation of the G327.3 continuum and dense gas emission
  data. The left and central panels present in color the 1.3\,mm
  continuum Stokes $I$ and linearly polarized $P$ emission,
  respectively. The contours start at the 3$\sigma$ levels continuing
  in $6\sigma$ steps up to 42$\sigma$, then increasing further in
  84$\sigma$ steps ($1\sigma$ are 1.5\,mJy\,beam$^{-1}$ and
  50\,$\mu$Jy\,beam$^{-1}$ for Stokes $I$ and linearly polarized
  emission, respectively). The central core (C) and two positions
  toward the north-east (NE1 \& NE2) are marked. The box in the middle
  panel shows the zoom-region presented in the right panel. There, the
  color-scale presents the 1st moment (intensity-weighted peak
  velocity) of $^{13}$CH$_3$CN$(12_4-11_4)$. The constant-length line
  segments show in all three panels the magnetic field orientation
  (polarization angles rotated by 90\,deg) derived from the linearly
  polarized continuum data above the $2\sigma$ level (independent of
  the polarization fraction). Linear scale bars are presented in the
  left and right panels, the synthesized beam ($1.16''\times 0.96''$)
  is shown in all panels in the bottom-left corner.  The green circle
  in the right panel outlines the $2''$ radius aperture where rotation
  appears to distort the magnetic field (see also section
  \ref{quantitative}).}
\label{pol_all}
\end{figure*}

Here, we are presenting mm-wavelength polarization observations with
the Atacama Large Millimeter Array (ALMA) toward an active high-mass
star-forming region, the bona-fide massive hot core G327.3. This
region is at a distance of $\sim$3.1\,kpc and has a luminosity of
$\sim 10^5$\,L$_{\odot}$ and a mass reservoir of
$\sim$950\,M$_{\odot}$ \citep{caswell1995,wyrowski2006}. It hosts
CH$_3$OH class II maser emission and a line rich, star-forming hot
molecular core (T$\geq$100\,K,
\citealt{walsh1998,wyrowski2006,leurini2013}) where highly excited
NH$_3$(5,5) data reveal a rotating central structure
\citep{beuther2009c}. The main question we address is whether gravity,
rotation and/or magnetic fields are dominating the dynamics in this
region.

\section{Observations}
 
The hot core G327.3 was observed as a cycle 6 program (id
2018.1.01449.S). The three observing blocks were observed on March 25,
2019, with total observing times of roughly 101, 99 and 112\,min,
respectively. In total, 41 to 43 effective antennas were in the array,
covering baselines between 14 and 331\,m. The on-source time for
G327.3 was roughly 40\,min per scheduling block, hence $\sim$2\,h
on-source time in total. 

As phase calibrator we used the quasar J1603-4904, the polarization
calibration was conducted with J1550+0527, and bandpass and flux
calibrations were done with J1427-4206. Calibration was done with the
CASA pipeline version 5.6.1 following the ALMA provided calibration
scripts. The phase center of G327.3 was R.A.~(J2000.0)=15:53:07.72 and
Decl.~(J2000.0)=$-$54:37:06.1 while the rest velocity of the source is
$v_{\rm{lsr}}=-46.0$\,km\,s$^{-1}$. The source was observed in the
1.3\,mm band with four spectral windows centered at 230.852, 229.152,
216.422 and 214.535\,GHz. While the first window had a width and
spectral resolution of 0.9375\,GHz and 0.977\,MHz, respectively, the
other three spectral windows had a width and resolution of 1.874\,GHz
and 1.953\,MHz, respectively. Since these are high-sensitivity data
and the region is a line-rich hot core, there is essentially no
line-free continuum part in the data. This is clearly the case for the
Stokes $I$ data, but Stokes Q and U are much weaker in line emission,
and one sees mainly the CO(2--1) spectral line in those two
polarization datasets. For all Stokes products (I, Q and U), we
collapsed the whole bandpass but excluding the CO(2--1) spectral
line. While that results in an overestimation of the Stokes $I$
continuum emission, the linearly polarized Q and U datasets should
represent the real polarized continuum well without significant line
contamination. Because of the line contamination of the Stokes I data,
we refrain from showing a polarization fraction map.

Self-calibration was applied within CASA improving especially the rms
in the Stokes I continuum image by more than a factor 3.  We imaged
the data with the \textsf{tclean} task in CASA experimenting with
different robust parameters. To optimize for the sensitivity of the
polarized emission in Stokes Q and U, the final data products
presented here were imaged with a robust parameter of 0.5, resulting
in a spatial resolution of $1.16''\times 0.96''$ (PA.~81.4\,deg).  We
de-biased the polarization data by estimating the rms $\sigma_P$ of
the linearly polarized data $P$ as
$\sigma_P=\sqrt{(Q\times\sigma_Q)^2+(U\times\sigma_U)^2)/(Q^2+U^2)}$,
where $\sigma_Q$ and $\sigma_U$ are the rms values of the Stokes Q and
U images \citep{dallolio2019}. The final $1\sigma$ rms of the total
intensity Stokes $I$ and linearly polarized
$P=\sqrt{Q^2+U^2-\sigma_P^2}$ images are 1.5\,mJy\,beam$^{-1}$ and
5\,$\mu$Jy\,beam$^{-1}$, respectively. 

We note that for linearly polarized emission, ALMA is not sensitive to
very large scales. While ALMA officially only guarantees reliable
polarization data within the inner 1/3 of the primary beam
(corresponding to the inner $\sim 9''$), recent analysis of nearby
pointings in polarization data revealed that they can be combined also
over more extended scales within the primary beam and for
mosaics \citep{beuther2018,hull2020}. Therefore, with the given high
signal-to-noise ratio the polarized emission and the associated
polarization angles are trustworthy out to at least $10''$ from the
field center.

While the polarized spectral line emission as well as the entire
Stokes $I$ spectral line cube will be analyzed in forthcoming
publications, for the spectral lines, we are presenting
in Fig.~\ref{pol_all} (right panel) the kinematic properties of the
dense gas traced by $^{13}$CH$_3$CN$(12_4-11_4)$ with a rest frequency
and upper level energy of 214.310\,GHz and $E_u/k_b$=181\,K,
respectively. The data were imaged with the (almost) native spectral
resolution of 3\,km\,s$^{-1}$. The $1\sigma$ rms value within each
channel is roughly 7.3\,mJy\,beam$^{-1}$.

To get a feeling about the optical depth of the dust continuum
emission and whether scattering may contribute to the polarization, we
converted the Stokes I image in the Rayleigh-Jeans limit to brightness
temperatures. While the central peak position reaches a brightness
temperature of $\sim$44\,K, the mean temperature of the entire central
core is $\sim$3.1\,K. Considering that this hot core region has gas
temperatures typically exceeding 100\,K \citep{beuther2009c}, the bulk
of the emission is optically thin. While the very central pixel may
have a small contribution from scattering, most of the polarization
emission should stem from dust grains gyrating around a rotation axis
that is aligned with the local magnetic fields (e.g.,
\citealt{lazarian2007}).

\section{Simulations}
\label{sim}

The comparison simulation was carried out by basically merging and
simplifying two recent modeling
scenarios \citep{koelligan2018,kuiper2018}. The model describes the
gravitational collapse of a magnetized pre-stellar core of 100 solar
masses of gas and dust within a sphere of 0.1\,pc in radius.  The
radial slope of the gas mass density is chosen to be proportional to
$r^{-1.5}$.  The core is initially in solid-body rotation with about
4\% of rotational to gravitational energy, turbulent motions are
neglected.  The initial temperature of the core is set to 10~K.  The
initial magnetic field strength is uniform in space and the magnetic
field is initially threading the pre-stellar core in a direction
parallel to the rotation axis in the weak-field regime with a
mass-to-flux ratio of 20 times the critical value  \citep{crutcher2012}
allowing the gravitational collapse to directly begin at the start of
the simulations.

We compute the temporal evolution of the system utilizing the most
recent version of our self-gravity radiation magneto-hydrodynamics
framework. For the magneto-hydrodynamics (MHD) equations, we make use
of the open source MHD code \emph{Pluto} \citep{mignone2007} version
4.1. We utilize a state-of-the-art constraint transport MHD solver
including the effects of ohmic dissipation for non-ideal/resistive
MHD.  For the dissipation strength, we follow  \citet{machida2007}, but
neglect the dependence on gas temperature.  The reconstruction is set
to be 2nd order accurate in time and space.

At the center of the infalling core, the formation and evolution of a
single star is computed via a sub-grid module.  The accretion history
of the protostar is given by the gas flow out of the computational
domain into a sink cell at the inner radial boundary of the
computational domain in spherical coordinates.  The current
gravitational mass of the protostar is given by the time integral of
the accretion history. For modeling the dust and gas continuum
radiation transport, we use a so-called two-temperature flux-limited
diffusion approximation in the linearization
approach \citep{commercon2011b}. The dust-to-gas mass ratio is fixed to
1\% throughout the evolution of the system.  Stellar radiative
feedback and radiation forces from the continuum radiation are
neglected in this simulation for simplicity: Radiation forces only
have a minor impact for the early phase of protostellar evolution
modeled here ($M_\star \le 20 \mbox{ M}_\odot$) and the heating effect
during the protostellar evolution does not modify the large-scale
magnetic field morphology studied herein. Self-gravity of the gas is
included in the numerical model via a diffusion Ansatz for solving the
Poisson equation \citep{kuiper2010}. Angular momentum and mass transport
by gravitational torques is modeled in the axially symmetric accretion
disk via a sub-grid module for
alpha-shear-viscosity \citep{mignone2007,kuiper2010}.

The simulation is carried out on a two-dimensional grid in spherical
coordinates assuming axial and midplane symmetry.  The dimension of
the computational domain in the radial direction extends from 3\,AU up
to 0.1\,pc. The dimension of the computational domain in the polar
direction extends from 0\,deg at the rotation/symmetry axis down to
90\,deg at the disk’s midplane.  To recover the basic morphology of
gravitational infall on large scales and disk formation on smaller
scales, a fairly low spatial resolution is required: The computational
domain consists of 56 grid cells in the radial direction and 10 grid
cells in the polar direction; the grid resolution in the radial
direction increases linearly with radius toward the origin of the
computation domain; the width of the grid cells in the radial
direction is identical to the width in the polar direction.  At the
inner radial boundary, we adopt a semi-permeable wall boundary
condition allowing fluxes out of the computational domain (mimicking
accretion onto the central protostar), but inhibiting fluxes into the
computational domain.  At the outer radial boundary, we adopt a
semi-permeable wall boundary condition as well allowing fluxes out of
the computational domain, but inhibiting fluxes into the computational
domain.  Zero-gradient boundary conditions are applied at the
boundaries in the radial direction for gas pressure and magnetic
field, the temperature of the radiation field is fixed to 10 K.  In
the polar direction, axially and midplane symmetric boundary
conditions are applied at the upper and lower end of the computational
domain.

The initially super-critical pre-stellar core collapses under its own
gravity.  In the early evolution, the system is dominated by gravity,
and hence, leads to radial infall.  A high-mass protostar is formed at
the center of the infalling core.  Later in the evolution, a
circumstellar disk forms around the protostar, and the disk grows with
time. A magnetically-driven, collimated, high-velocity
($>100$\,km\,s$^{-1}$) jet is launched into the bipolar direction
perpendicular to the forming disk. At the simulation time of 30\,kyrs
after the start of collapse a 17\,M$_{\odot}$ protostar has formed at
the center. In the subsequent evolution the protostellar mass and size
of the rotating disk continue to grow.

\section{Results}

\subsection{Morphologies and kinematics}

The main observational results from these polarization observations
are presented in Figure \ref{pol_all}. The 1.3\,mm dust continuum
Stokes $I$ total intensity is dominated by a massive dense core at the
center. Furthermore, we identify filamentary structures that lead
toward the central core from the south-western and north-eastern
direction.  The linearly polarized emission $P$, shown in the middle
panel of Fig.~\ref{pol_all}, exhibits emission at similar locations to
the Stokes $I$, i.e., strong emission in the center and an extension
toward the north-east where also the main Stokes $I$ filament is seen.

Even more important are the position angles of the linearly polarized
emission. Assuming that the polarized emission is produced by dust
grains gyrating around a rotation axis that is aligned with the local
magnetic fields (e.g., \citealt{lazarian2007}), we rotated all angles
by 90\,deg and show these rotated angles outlining the direction of
the magnetic field in all panels of Fig.~\ref{pol_all}. The morphology
of the derived magnetic field structure is neither uniform nor does it
clearly resemble an hourglass-like structure. However, it is
intriguing that this magnetic field structure is oriented radially
toward the center of the main Stokes $I$ emission peak from almost all
azimuthal directions.  This indicates that the magnetic field is
dragged toward the gravitational center of this active star-forming
hot molecular core.

Examining the filamentary emission oriented to the north-east of the
center (toward NE1 and NE2 in Fig.~\ref{pol_all}), the magnetic field
is bent in a U-like shape, also directed radially toward the center of
the main mm continuum peak. This U-like shape of the magnetic field is
predicted by magneto-hydrodynamic simulations of cloud collapse that
form filamentary structures along which the gas is fed toward the
gravitational center \citep{gomez2018}. We will come back to this
directional change of the magnetic field in section \ref{discussion}.

Zooming toward the center, we investigated the spectral line
properties of the dense gas tracer methyl cyanide (specifically the
$^{13}$CH$_3$CN$(12_4-11_4)$ transition). The right panel of
Fig.~\ref{pol_all} shows the first moment map (intensity-weighted peak
velocities) of the dense gas, in which one sees a velocity gradient
from red- to blue-shifted from the north-west toward the south
east. This velocity gradient may be caused by rotation of the dense
central core. Toward the center of this core, the magnetic field
morphology is not directed radially toward the center anymore but it
transforms into a potentially rotating structure.  Multiplicity and
other potential substructure below our resolution limit
($\sim$3000\,AU) may influence the kinematic and polarization
observations. However, even a multiple system is likely to undergo
rotation and is hence likely to not change the conclusion that
rotation influences the magnetic field morphology at the center of our
observations. More detailed multiplicity analysis can only be
conducted with higher-resolution follow-up observations.

An important aspect is the inclination angle at which we are observing
the system. Carbon monoxide CO(2--1) data (from the same observing
program but not shown here) reveal high-velocity gas ($\Delta
v\geq$50\,km\,s$^{-1}$ from the velocity of rest), and the red- and
blue-shifted emission peaks are separated by only $\sim$1$''$
($\sim$3000\,AU). This indicates that we are looking almost face-on
into a rotating and collapsing system where the outflow is roughly
along the line of sight. In such a face-on orientation one would not
expect an hourglass-shaped magnetic field distribution, but the
morphology should just appear as dragged radially toward the
  center, similar to what we observe \citep{frau2011}. Therefore, the
data do not allow us to reject the possibility that an hourglass-like
morphology may be present also in G327.3, just masked by an almost
pole-on observing configuration.

\subsection{Quantitative analysis}

\subsubsection{Masses, column densities  and magnetic field strengths}
\label{quantitative}

To get an estimate about the column densities and masses in this
region, we derived the peak flux densities and the integrated fluxes
within the $3\sigma$ contours for the main central contiguous
structure and the two filamentary extensions in north-eastern
direction (labeled in Fig.~\ref{pol_all}, left panel as C, NE1,
NE2). Assuming optically thin dust continuum emission at an average
temperature of this typical hot core of 100\,K \citep{beuther2009c}, a
gas-to-dust mass ratio of 150 \citep{draine2011} and using a dust
absorption coefficient $\kappa = 1.11$\,cm$^2$g$^{-1}$ at densities of
$10^8$\,cm$^{-3}$ \citep{ossenkopf1994}, the central gravitational
attractor contains roughly 400\,M$_{\odot}$ whereas the two
filamentary sub-structures are far less massive around $\sim$2.7 and
$\sim$1.5\,M$_{\odot}$. Assuming a temperature uncertainty of a factor
2, the masses can vary also approximately by a factor 2. More details
are given in Table \ref{masses}. Adding a factor 2 uncertainty for the
dust absorption coefficient, the mass uncertainty can be as high as a
factor 4. These numbers should only be considered as rough estimates,
and more in-depth analysis considering the temperature structure and
potential spatial filtering will be conducted in a follow-up analysis
including also the spectral line data.

  \begin{table}[htb]
    \caption{Stokes $I$ continuum parameters}
    \label{masses}
    \begin{tabular}{lrrrr}
      \hline \hline
      Source & $S$ & $S_{\rm peak}$ & $M$ & $N$ \\
      & (Jy) & (Jy\,beam$^{-1}$) & (M$_{\odot}$) & ($10^{24}$cm$^{-2}$)\\
      \hline
      C   & 9.81  & 1.966 & $435^{+485}_{-223}$  & $13.4^{+15.0}_{-6.9}$ \\
      NE1 & 0.065 & 0.046 & $9.2^{+10.3}_{-4.7}$ & $0.4^{+0.5}_{-0.2}$ \\
      NE2 & 0.035 & 0.031 & $4.0^{+4.4}_{-2.1}$  & $0.2^{+0.3}_{-0.1}$ \\
      \hline \hline  
    \end{tabular}
    ~\\ Notes: Sources are labeled in Fig.~\ref{pol_all}. Flux
    densities $S$, peak intensities $S_{\rm peak}$, masses $M$ and gas
    column densities $N$ are given.  The error margins for $M$ and $N$
    correspond to an uncertainty of a factor 2 in the temperature.
  \end{table}

Estimates of the magnetic field strength via the
Davis-Chandrasekhar-Fermi (DCF) method
\citep{davis1951,chandrasekhar1953} are unfortunately barely feasible
for this region. One of the main assumptions of the DCF method is that
the dispersion of angles is the result of transverse incompressible
Alfven waves, and that the dispersion of polarization angles is
relatively small ($<$25\,deg, \citealt{ostriker2001}). Since this is
obviously not given in the region investigated here, the
Davis-Chandrasekhar-Fermi method is not properly
applicable. Furthermore, the derived structure function is almost flat
at an angle value of $\sim$52\,deg at scales beyond $\sim$1$''$, which
is consistent with a random distribution. As this scale is only
marginally larger than the Nyquist-sampled beam, estimates of, e.g.,
the turbulent-to-mean-field ratio using the second-order structure
(e.g., \citealt{hildebrand2009,houde2009,koch2010}) would not be
reliable in this region. As we have also shown in our analysis, the
magnetic field is not randomly oriented but preferentially
perpendicular to the isophote Stokes $I$ contours.

We can approximately quantify what velocities are required if the
rotational inner structure is centrifugally supported. The region
where the magnetic field appears rotationally distorted is about $2''$
in radius from the center. At 3.1\,kpc distance that corresponds to a
radius of 6200\,AU. Measuring the mass with the above assumptions from
the dust continuum data only within that aperture (Fig.~\ref{pol_all},
right panel), we find approximately 331\,M$_{\odot}$. Assuming
equilibrium between the centrifugal and gravitational forces at the
outer radius $r$ of the disk, the velocity $v$ corresponding to the
enclosed dynamical mass $M$ can be estimated via (with the
gravitational constant $G$)

\[
v = \sqrt{\frac{MG}{r}}
\]
\[
\Rightarrow v\rm{[km\,s^{-1}]} =
\sqrt{\frac{\it{M}\rm{[M_{\odot}]}}{1.13\times 10^{-3}\,\it{r}\rm{[AU]}}} 
\]
\[
\Rightarrow v = \sqrt{\frac{331}{1.13\times 10^{-3} \times 6200}}\rm{km\,s^{-1}} = 6.9\,km\,s^{-1}
\]
    
How do these 6.9\,km\,s$^{-1}$ compare to our observations? Within the
given area of $4''$ diameter, the maximum velocity difference is about
2\,km\,s$^{-1}$, or over the radius $r$ about 1\,km\,s$^{-1}.$ To
account for the projection in the observations, we need to multiply
the above equation with $sin(i)$ where $i$ is the inclination angle
and 0 corresponds to a face-on geometry. With $i\sim 9$\,deg, the
required velocity $v$ is consistent with our observed velocities. This
confirms our previous assessment that we are observing the system with
a geometry where the rotating structure has to be almost face-on.

In principle, not just projection but also magnetic breaking can
reduce the observed velocities, and these two parameters are
degenerate. Hence, we cannot properly differentiate whether the actual
inclination angle $i$ is indeed only $\sim$9\,deg without magnetic
breaking, or whether the angle is somehow larger and magnetic breaking
may contribute to further velocity reduction. 

\subsubsection{Histogram of Relative Orientations (HRO)}

Since we cannot properly estimate the magnetic field strength, a
quantitative comparison of the corresponding energy terms is not
possible. However, we can quantify the morphological and kinematic
results by investigating the relative orientation between the magnetic
field direction and the isophote contours from the Stokes $I$
continuum emission (Fig.~\ref{pol_all} left panel). This is done
quantitatively by means of the histogram of relative orientations
(HRO) introduced for the Planck magnetic field studies
\citep{soler2013,soler2016,soler2019b}.

\begin{figure}[htb]
\includegraphics[width=0.49\textwidth]{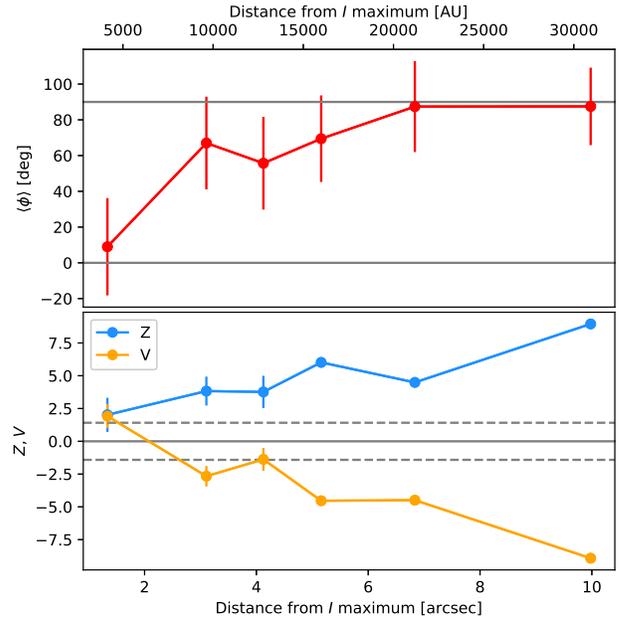}
\caption{Histogram of relative orientations (HRO) analysis between the
  magnetic field and the orientation of isophote contours of the
  Stokes $I$ data with respect to the azimuthally averaged
  distance from the core center. The (projected) Rayleigh statistics
  $Z$ and $V$ in the bottom panel describe the degree of correlation
  (see main text). The dashed lines correspond to what is expected for
  a random distribution of angles, $\pm\sqrt{2}$. The top panel
  presents the relative angle distribution where 90\,deg would mean
  that magnetic field and intensity contours are perpendicular. The
  plots are derived with equal number of pixels in each bin 
    starting at a $5\sigma$ threshold in Stokes I and the polarized
    intensities.}
  \label{hro} 
\end{figure}

The output of the HRO is the distribution of the relative orientation
angles ($\phi$) between the structures in Stokes $I$ and the
magnetic field. This distribution is characterized by the mean
orientation angle $\left<\phi\right>$ and two statistical tests from
circular statistics, the Rayleigh test $Z$ and the projected Rayleigh
test $V$ \citep{brazier1994,jow2018}.  The values of $Z$ can be
understood as the total displacement from the center of a 2D plane
resulting from unitary steps in the orientations defined by
$\phi$, thus, $Z$\,$\approx$\,0 corresponds to a random
distribution of angles. The values of $V$ are the projection of that
total displacement in the directions of 0 and 90 deg, represented by
$V>0$ and $V<0$. Consequently, positive or negative $V$-values indicate
that the angles between Stokes $I$ and the magnetic field are mostly
parallel or mostly perpendicular, respectively. Figure \ref{hro}
presents the corresponding parameters as a function of the distance
from the peak position.

The values of the projected Rayleigh statistic indicate that for most
of the distance bins, the distribution is unimodal and consistent with
a preferential orientation of the magnetic field perpendicular to
  the Stokes I isophote contours significantly above from what is
expected for a random distribution of angles, $\pm\sqrt{2}$ (dashed
lines in Fig.~\ref{hro}).  The mean angle orientation
$\left<\phi\right>$ indicates that the mean direction is close to 90
deg, within the confidence limits. Looking at the spatial structure in
more detail, for the large distance bins the projected Rayleigh
  statistic $V$ is consistent with the magnetic field being oriented
  almost perpendicular to the isophote contour levels (Figs.~\ref{hro}
  and \ref{pol_all}). Getting close to the center, the orientation
between magnetic field and contours becomes less perpendicular and
again consistent with a random distribution. This inner change in
orientation can also be seen on the smallest scales in
Fig.~\ref{pol_all} (right panel) and is likely attributed to rotation
of the inner region becoming more important. The same result can
be derived from the potentially more intuitive relative angle
distribution $\phi$ (Fig.~\ref{hro}): far outside, one finds
large angles between the magnetic field and the isophote
contours. Getting closer to the center, the angle distribution gets
smaller and approaches 0 as expected if centrifugal forces
progressively counteract the central gravitational pull.  

\begin{figure}[htb]
\includegraphics[width=0.49\textwidth]{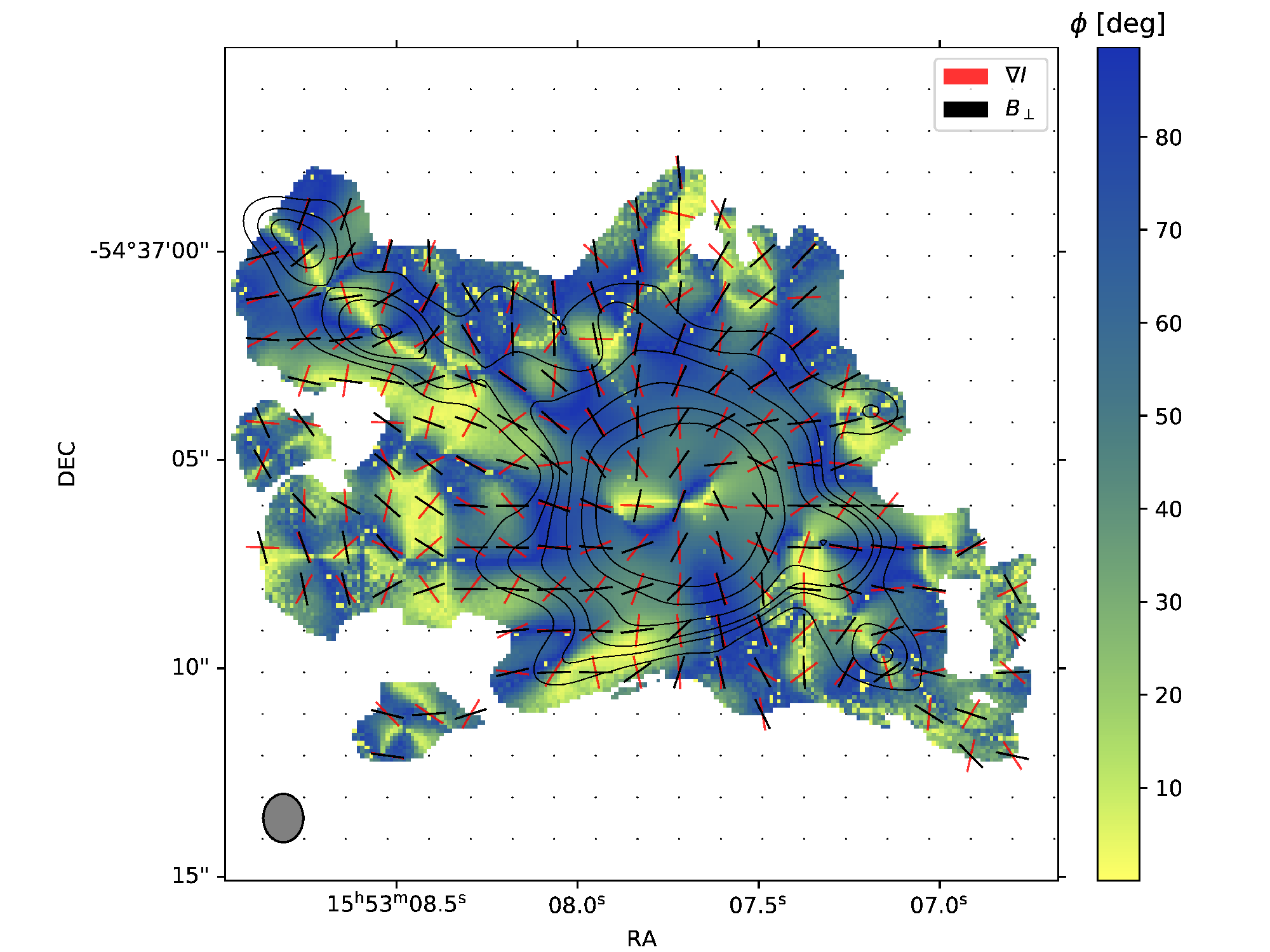}
\caption{Angle distribution between Stokes I isophote contours and
  magnetic field orientation. Blue colors or large angles $\phi$
  correspond to the magnetic field and isophote contours almost
  perpendicular to each other whereas yellow and low angle $\phi$
  values correspond to largely parallel. The black line segments
  correspond to the magnetic field orientation, and the red segments
  to the orientation of the Stokes I intensity gradient (that is
  perpendicular to the isophote contours). The contours are the Stokes
  $I$ map with contour levels of 0.01, 0.02, 0.03, 0.06, 0.125, and
  0.25\,Jy\,beam$^{-1}$. The beam is shown at the bottom-left, and the
  dots correspond to the grid where the line segments are plotted at.}
  \label{phi_map} 
\end{figure}

A different way to visualize the results is by plotting the angle
$\phi$ between isophote contours and magnetic field orientation in a
2-dimensional map, similar to the sin$(\omega)$ maps presented in
\citet{koch2018}. We prefer to present the real angles instead of the
sin of the angles because the latter is slightly skewed to larger
values due to the sin-function. But qualitatively, the two measures
represent the same. The corresponding $\phi$ map is shown in
Fig.~\ref{phi_map}. The blue features in nearly a ring-like structure
around the central peak position clearly show the perpendicular nature
of the magnetic field to the isophote contours at almost all angles
around the central core, confirming the above interpretation. Yellow
channel-like features of low angles or nearly parallel structures
between isophote contours and magnetic field are mainly found at the
spine of the north-eastern filament, and also, although weaker, in the
south-western filamentary structure. These yellow spines are
consistent with the U-like structures in the filament discussed
above. Furthermore, in the central region where rotation appears to
become more important, the angle distribution is less homogenous but
rather varies between large and small values, again confirming that
additional forces, such as centrifugal forces, come into play.

\section{Discussion}
\label{discussion}

The analysis of G327.3 investigates spatial scales between a few
thousand and several ten-thousand astronomical units. In this regime
the magnetic field is perpendicular to the gas isophote contours in
the outskirts of the core and changes orientation close to the
center. Setting that into context with results derived for larger
scales (0.4-40\,pc) molecular cloud data from Planck and Herschel
observations, it is found that at very low densities (typically on
scales on the order of $\sim$10\,pc) the magnetic field is oriented
mostly parallel to the gas structure, and that at higher column
densities (typically scales of $\sim$1\,pc), the orientation changes
to almost perpendicular to the gas structure
\citep{soler2013,soler2019b}. These Planck results are typically
interpreted in the strong field regime where the gas flow follows the
magnetic field orientation. Interestingly, our data with the magnetic
field structures perpendicular to dense gas on scales of tens of
thousands of AU exhibit a similar observational correlation, just on
very different scales and also because of different physical
processes.


\begin{figure*}[htb]
\includegraphics[width=0.4\textwidth]{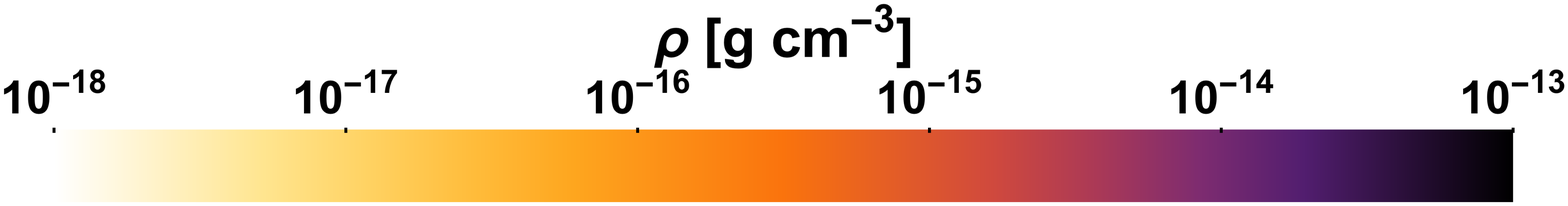}\\
\includegraphics[width=0.4\textwidth]{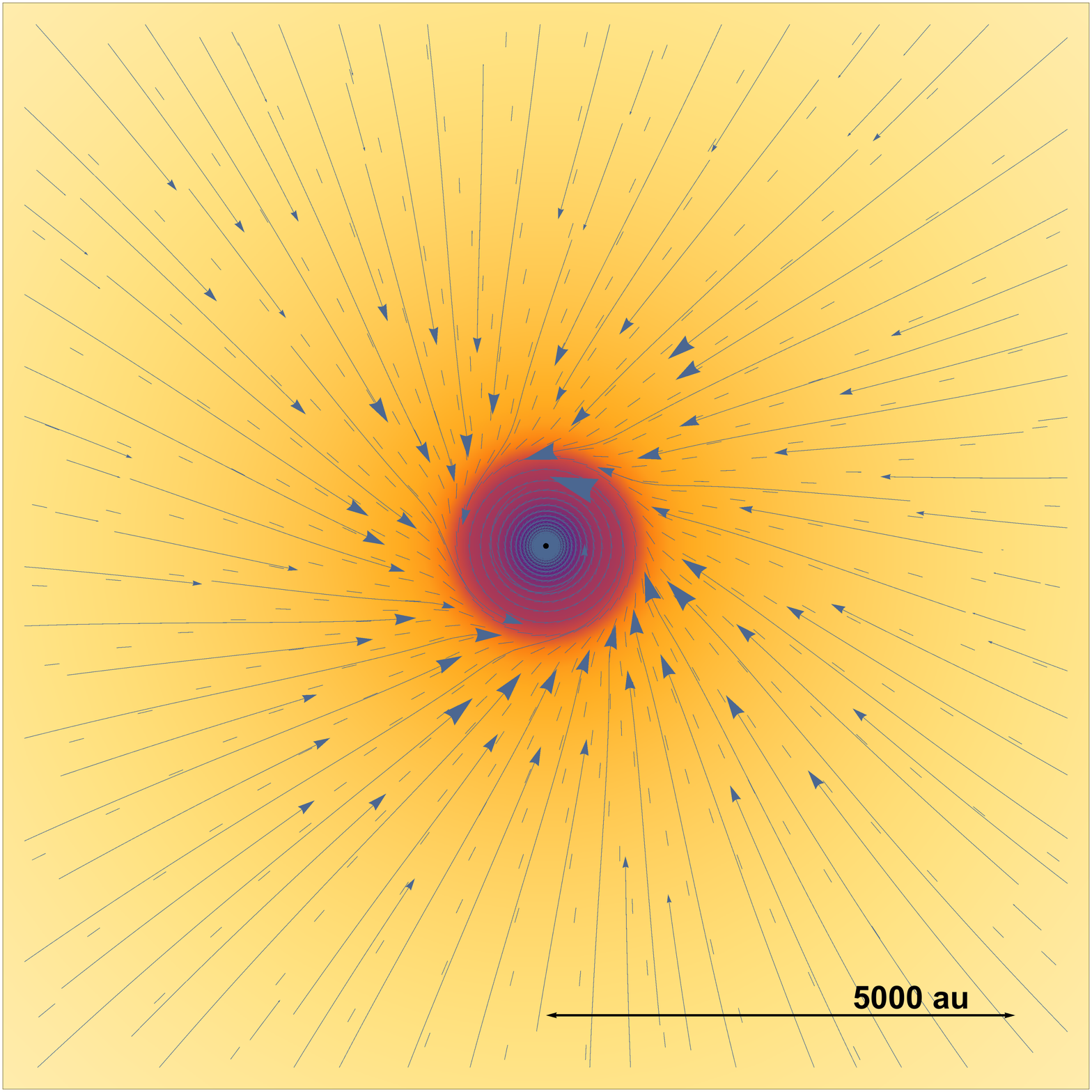}
\includegraphics[width=0.59\textwidth]{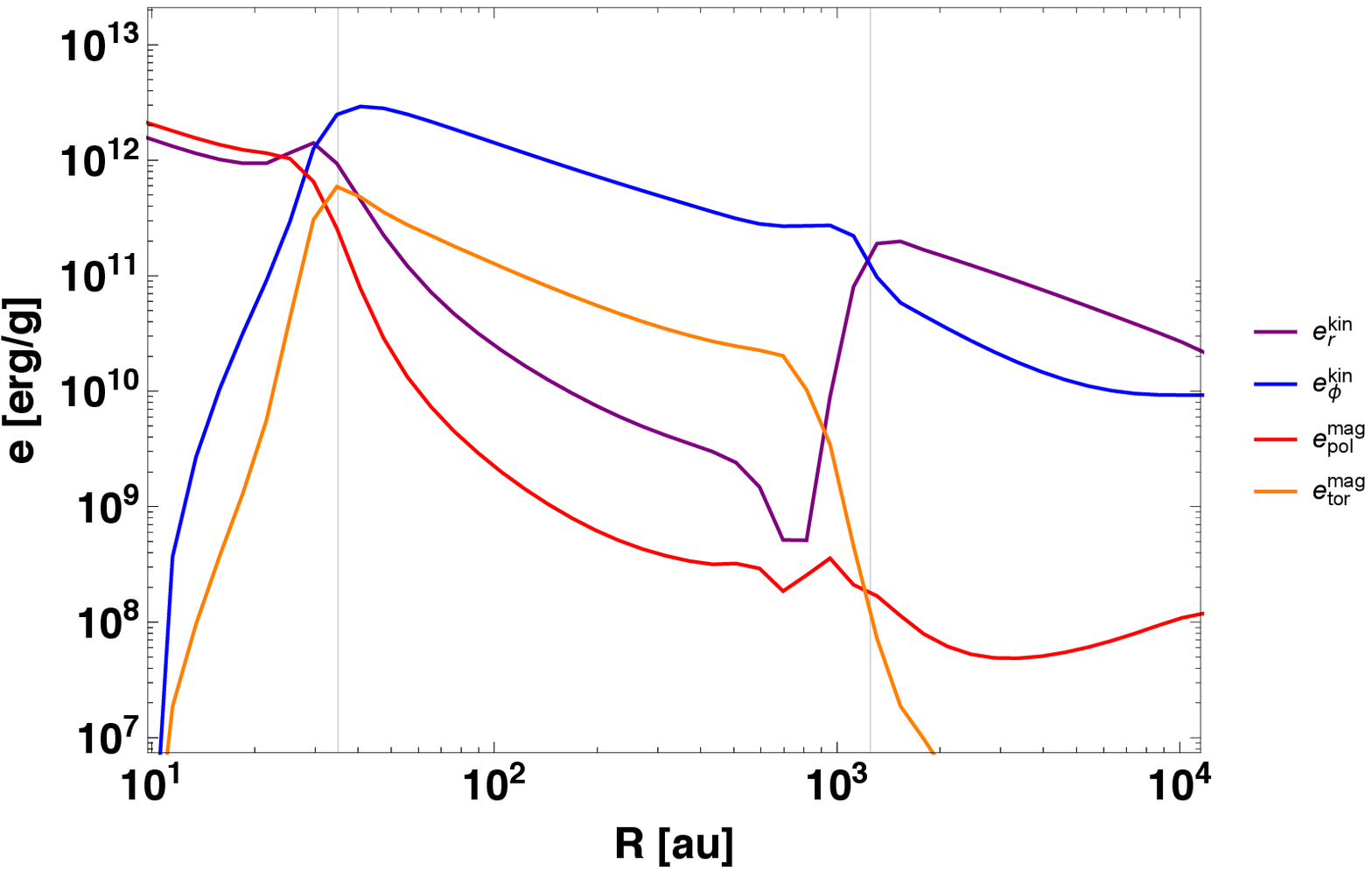}
\caption{Numerical cloud collapse model.  Face-on view of the gas mass
  density distribution (color) and magnetic field morphology (segments
  and arrows) in the midplane.  The left image shows a zoom onto the
  central region at 30\,kyr after onset of the gravitational
  collapse. Blue line segments represent local magnetic field
  orientation, the arrows connect them to trajectories/field
  lines. Details are given in Sec.~\ref{sim}. The right panel presents
  the radially averaged energy profiles for the radial/azimuthal
  specific kinetic energies ($e_{\rm r}^{\rm kin}, e_{\phi}^{\rm
    kin}$) and the poloidal/toroidal specific magnetic energies
  ($e_{\rm pol}^{\rm mag}, e_{\rm tor}^{\rm mag}$), respectively. The
  vertical lines at 35 and 1250\,AU mark approximate transition
  regions.}
  \label{model} 
\end{figure*} 

On the smallest scales another change of orientation between dense gas
and magnetic field apparently occurs. We propose that this change then
could be associated with the rotational properties of any potential
disk in the inner part of the star-forming region, also suggested for
a few other regions (e.g.,
\citealt{girart2013,hull2014,kwon2019}).

Figure \ref{model} presents qualitative and quantitative results from
magneto-hydrodynamic cloud collapse simulations where gravity
dominates over magnetic fields (see section \ref{sim} for details on
the simulations). These simulations show field morphologies from a
centrally directed pattern in the outer parts of the cloud core to a
more circular pattern in the inner disk regions, resembling closely
the observed morphology of the magnetic field (Fig.~\ref{pol_all}). On
large scales, radial specific kinetic energies $e_{\rm r}$ dominate
and the magnetic field is dragged into the core's central region due
to gravity-dominated infall. On the smaller scales of the disk
formation, centrifugal specific energies {$e_{\phi}^{\rm kin}$ become
  important and the magnetic field topology is transformed in a
  toroidal structure (increase of $e_{\rm tor}^{\rm mag}$)}.  In the
very inner region of the simulations below 35\,AU radius (unresolved
by our current observations), the energies are dominated by the
specific poloidal magnetic component $e_{\rm pol}^{\rm mag}$ that is
also responsible for removing angular momentum (strong decrease of
$e_{\phi}^{\rm kin}$) and driving an outflow. The inner regions (below
radii of 1000\,AU) will also be subject to future even
higher-spatial-resolution observations with ALMA.

Comparing our results, in particular the $\phi$ map
(Fig.~\ref{phi_map}), with the angle distributions found previously by
\citet{koch2018}, we find similarities but also differences. While
\citet{koch2018} also observe large angles around the central cores
with magnetic field to isophote contours in preferred perpendicular
orientation, they find toward several cores so-called ``yellow
channels'' with almost parallel structures between magnetic field and
isophote contours. Such latter ``yellow channels'' we only find toward
the filamentary structures in the northeast and southwest but not
around the central massive core. This difference may be due to large
degree to the special geometry of G327. We are observing the region
almost face-on where the mean field orientation has a significant
contribution along the line of sight. In that geometry the circular
symmetric structures between magnetic field and isophote contours are
expected. In contrast to that, in more edge-on like orientations, gas
is expected to be fed toward the center in more channel-like
structures (e.g., Fig.~11 in \citealt{koch2013}). In ideal
magneto-hydrodynamics the magnetic field is dragged by the flow
because the core at the protostellar stage has to be supercritical
(otherwise a star would not form). In a face-on orientation, even if
the core is strongly oblate, most of the accretion occurs through the
equatorial plane and therefore perpendicularly to the field
lines. Furthermore, G327 is still in a young evolutionary stage with
strong interaction between the central core and the environment. For
example, feeding of the core by the surrounding filamentary structures
can further distort the field geometry.

Based on an analytic and numeric analysis of the conditions
  in ideal magnetohydrodynamic turbulence, it was concluded that the
change in orientation may be associated with convergent gas flows
and/or gravitational collapse \citep{soler2017}.  The U-shaped change in
magnetic field orientation in the north-eastern filament is also
reminiscent of a converging gas flow in the sense that the gas flows
first onto the filament and then along the filament toward the main
gravitational well \citep{gomez2018}.

\section{Conclusions}

Combining the results of the centrally directed magnetic field around
the central main core, the U-like magnetic field shape in the
filamentary extensions, and the inner morphology change indicative of
rotation, all this strongly indicates that gravity and centrifugal
forces drag the magnetic field along during the collapse of this
high-mass star-forming region. In the outer region, the gas and
magnetic field follows the filament, then collapses toward the center
and there transforms into a rotational structure which potentially
feeds an inner still unresolved accretion disk. This hot core region
is in an evolutionary stage with ongoing active star formation, and
gravity has to be the dominating force in this system. Our comparison
with supercritical mass-to-flux simulations is suggestive for a
weak-field scenario for G327.3. However, the pole-on orientation does
not allow us to exclude an hourglass-like morphology with strong-field
initial conditions. Other studies also show opposing results whether
weak (e.g., \citealt{beuther2010c,girart2013}) or strong fields (e.g.,
\citealt{hull2014,li2015,pillai2015}) are more typical. Future sample
studies of different evolutionary stages as well as regions with
different inclination angles are needed to further constrain the
initial magnetic field conditions and the evolutionary changes.

\acknowledgments

We like to thank the referee for the insightful comments improving the
paper. This paper makes use of the following ALMA data:
ADS/JAO.ALMA\#2018.1.01449.S. ALMA is a partnership of ESO
(representing its member states), NSF (USA) and NINS (Japan), together
with NRC (Canada), MOST and ASIAA (Taiwan), and KASI (Republic of
Korea), in cooperation with the Republic of Chile. The Joint ALMA
Observatory is operated by ESO, AUI/NRAO and NAOJ.  HB acknowledges
support from the European Research Council under the Horizon 2020
Framework Program via the ERC Consolidator Grant CSF-648505. HB also
acknowledges support from the Deutsche Forschungsgemeinschaft via SFB
881, The Milky Way System (sub-project B1). RK acknowledges financial
support via the Emmy Noether Research Group on Accretion Flows and
Feedback in Realistic Models of Massive Star Formation funded by the
German Research Foundation (DFG) under grant no. KU 2849/3-1 and KU
2849/3-2.



\end{document}